\documentclass{moriond}

\def\al{\alpha}

\def\be{\begin{equation}}
\def\ee{\end{equation}}
\def\bea{\begin{eqnarray}}
\def\eea{\end{eqnarray}}

\newcommand{\Photo}{}

\usepackage{xcolor,braket,amsmath}

\newcommand{\refeq}[1]{eq.~(\ref{eq:#1})}
\newcommand{\refsec}[1]{sec.~\ref{sec:#1}}

\newcommand{\reffig}[1]{fig.~\ref{fig:#1}}
\newcommand{\FM}[2][BK^{(*)}]{\ensuremath{\mathcal{F}^{#1}_{#2}}}
\newcommand{\HM}[2][BK^{(*)}]{\ensuremath{\mathcal{H}^{#1}_{#2}}}
\newcommand{\VM}[2][BK^{(*)}]{\ensuremath{\mathcal{V}^{#1}_{#2}}}
\newcommand{\SP}[2]{\ensuremath{\mathcal{S}_{#1}^{#2}}}

\begin{document}
\vspace*{4cm}
\title{Theoretical predictions for $b\to s \mu^+ \mu^-$ decays}

\author{ N. Gubernari }

\address{
    DAMTP, University of Cambridge, Wilberforce Road,\\
    Cambridge, CB3 0WA, United Kingdom
}

\maketitle
\abstracts{
I review the state of the art of the theoretical calculations for decays mediated by $b\to s \ell^+ \ell^-$ transitions, for $\ell=e,\mu$.
I focus on the predictions of observables in $B\to K \mu^+\mu^-$,  $B\to K^* \mu^+\mu^-$, and  $B_s\to \phi \mu^+\mu^-$ decays, as many of these predictions are in tension with the corresponding experimental measurements.
I also briefly discuss the $\Lambda_b\to \Lambda \mu^+\mu^-$ decay and present a new calculation for this channel.
Special emphasis is placed on the non-local contributions, as they are the largest systematic uncertainties in these decays.
The current theoretical calculations for $b\to s \mu^+ \mu^-$ decays are not able to explain the tensions with the experimental measurements.
}

\section{Introduction}

One of the main goals of flavour physics is to probe the Standard Model (SM) using indirect searches.
To this end, very precise experimental measurements of flavour observables are compared with equally precise SM predictions.
This procedure provides strong constraints on New Physics (NP) and may eventually lead to a new discovery. 

The rare $B$ meson decays mediated by $b\to s \ell^+ \ell^-$ transitions are ideal for performing these indirect searches.
These decays have very small branching ratios in the SM ($\sim\!10^{-6}$) because they are loop mediated and GIM or CKM suppressed.
As a consequence, NP could generate a substantial contribution compared to the SM one.
For instance, in many $Z'$ and leptoquark models the $b\to s \ell^+ \ell^-$ transitions occur at tree-level and hence would only be suppressed by the mass of the new heavy mediator~\cite{Allanach:2023uxz,Crivellin:2023saq}.

Interestingly,  several measurements of observables in $B\to K \mu^+\mu^-$, $B\to K^* \mu^+\mu^-$, and $B_s\to \phi \mu^+\mu^-$ decays~\cite{LHCb:2014cxe,CMS:2024syx,LHCb:2016ykl,LHCb:2020lmf,LHCb:2021zwz} are in tension with the corresponding SM predictions~\cite{Parrott:2022zte,Gubernari:2022hxn}.
The observables that show tension are the differential branching ratios and the angular observables, which have been measured with moderate precision by both the LHCb and CMS experiments. 
The significance of the tensions depends on the process and the region of the phase space considered, and 9s in some cases above $4\sigma$~\cite{Parrott:2022zte}.
These tensions are commonly called the $b\to s \mu^+ \mu^-$ \emph{anomalies}.
Note that these anomalies form a coherent picture and can therefore be explained by the introduction of a new heavy mediator~\cite{Allanach:2023uxz,Crivellin:2023saq}.
It should be emphasised that these anomalies are not related to the previous anomalies in the lepton flavour universality ratios $R_K$ and $R_{K^*}$, which disappeared at the end of 2022~\cite{LHCb:2022qnv}.

On the one hand, the experimental measurements for the $b \to s \mu^+ \mu^-$ observables are considered reliable due to the background suppression and the efficient triggering of the muon tracks.%, since the backgrounds are reasonably well understood and small compared to the signal.
On the other hand, the theoretical predictions for $b \to s \mu^+ \mu^-$ decays are extremely challenging, due to the appearance of non-perturbative QCD effects.
Therefore, progress in the theoretical calculation is urgently needed to understand whether these anomalies are due to NP or to misestimated QCD contributions.

The rest of this document is structured as follows.
In \refsec{frame} I briefly review the theoretical framework by introducing the weak effective Hamiltonian and the form factors.
In \refsec{FFpred} I present the most recent local and non-local form factor calculations.
In \refsec{SMpred} I give the SM predictions and compare them with recent experimental measurements.
In \refsec{baryon} I discuss the importance of rare $\Lambda_b\to\Lambda \ell^+\ell^-$ decays before concluding in \refsec{conc}.

\section{Theoretical framework}
\label{sec:frame}

It is convenient to perform theoretical calculations for rare $b$-hadron decays in the framework of an effective field theory --- called the Weak Effective Theory (WET) --- rather than using the full SM Lagrangian. 
This allows to factorise the short- and long-distance contributions. 
The short-distance contributions are contained in the Wilson coefficients, and they are process independent.
Therefore, there is no need to recalculate the Wilson coefficients every time the WET is used, which is an enormous practical advantage.
In contrast, the long-distance contributions that are contained in the matrix elements of the effective operators are manifestly process dependent.
The WET Hamiltonian for $b \to s \ell^+ \ell^-$ transitions reads~\cite{Buchalla:1995vs}
\setlength{\abovedisplayskip}{8pt}
\setlength{\belowdisplayskip}{8pt}
\begin{align}
    H_{\rm eff}^{bs\ell\ell}
    = 
    -\frac{4 G_F}{\sqrt{2}} V_{tb} V_{ts}^*
    \sum_{i=1}^{10} C_i(\mu) {\cal O}_i(\mu) + \dots\,,
\end{align}
where the ellipsis indicates CKM-suppressed terms and the QCD and QED interaction terms are not explicitly shown.
The Wilson coefficients are denoted by $C_i$, while the effective operators are denoted by ${\cal O}_i$.
The scale is assumed to be $\mu\simeq m_b$.
For a very good approximation, it is sufficient to consider only the following operators:
\begin{align}
{\cal O}_1  & = (\bar{s}_L \gamma_\mu T^a c_L) (\bar{c}_L \gamma^\mu T^a
b_L)
\,, &
{\cal O}_2 & = (\bar{s}_L \gamma_\mu c_L) (\bar{c}_L \gamma^\mu b_L) \,,
&{\cal O}_7  = \frac{e}{16 \pi^2}
m_b (\bar{s}_L \sigma^{\mu\nu} b_R) F_{\mu\nu}
\,,  &
\nonumber\\
{\cal O}_9 & = \frac{e^2}{16 \pi^2}
(\bar{s}_L \gamma_\mu b_L) (\bar{\ell} \gamma^\mu \ell)
\,, &
{\cal O}_{10} & = \frac{e^2}{16 \pi^2}
(\bar{s}_L \gamma_\mu b_L) (\bar{\ell} \gamma^\mu \gamma_5 \ell)
\,.
\end{align}
The decay amplitude for $B\to K^{(*)} \ell^+\ell^-$ (and  $B_s\to \phi \ell^+\ell^-$) decays can be written as (neglecting QED corrections)
\begin{align}
\label{eq:amplitude}
{\cal A}(\bar B\to K^{(*)} \ell^+\ell^-) &=  {\cal N}
\bigg[ (C_9 L^\mu_{V} + C_{10} L^\mu_{A})\  \FM{\mu} \!
        -  \frac{L^\mu_{V}}{q^2} \Big\{  2 i m_b C_7\FM{T,\mu}\!  + 16\pi^2 \HM{\mu}\! \Big\}   \bigg]\ ,
\end{align}
where
\begin{equation}
\begin{aligned}
    {\cal N} &\equiv \frac{G_F\, \alpha\, V^*_{ts} V_{tb}}{\sqrt{2} \pi}
    \,,&
    L_{V(A)}^\mu &\equiv \bar u_\ell(q_1) \gamma^\mu(\gamma_5) v_\ell(q_2)
    \,,\\
    \label{eq:defME}
    \FM{\mu}\!(k, q) &\equiv \langle  K^{(*)}(k)|\bar{s}_L\gamma_\mu  b_L|\bar{B}(q+k)\rangle, &
    \FM{T,\mu}\!(k, q) &\equiv \langle  K^{(*)}(k)|\bar{s}_L \sigma_{\mu\nu} q^\nu  b_R|\bar{B}(q+k)\rangle, 
    \\
    \HM{\mu}\!(k, q) &\equiv
    i\!  \int d^4x\, e^{i q\cdot x}\,
    \bra {K^{(*)}(k)} T\big\{ j_\mu^{\rm em}(x), (C_1{\cal O}_1 + C_2{\cal O}_2)(0) \big\}  \ket{\bar B(q+k)}
    .  \hspace*{-30cm}&
\end{aligned}
\vspace*{-0.2cm}
\end{equation}
\vspace*{-0.1cm}As is readily apparent from these definitions, the $\FM{(T),\mu}$ are hadron-to-hadron matrix elements of local operators while the $\HM{\mu}$ are hadron-to-hadron matrix elements of non-local operators.
In general, these matrix elements can be written in terms of form factors (FFs):
\begin{align}
    &
    \FM{(T),\mu}(k, q) \propto \sum_\lambda  \FM{(T),\lambda}(q^2)\, \SP{\mu}{\lambda}(k, q)\ ,
    &&
    \HM{\mu}(k, q) \propto \sum_\lambda  \HM{\lambda}(q^2)\, \SP{\mu}{\lambda}(k, q)\ ,
    &
\end{align}
where the Lorentz structures $\SP{\mu}{\lambda}$ are known functions~\cite{Gubernari:2020eft}.
These equations follow from Lorentz invariance.

\section{Form factors predictions}
\label{sec:FFpred}

The FFs are dominated by long-distance contributions and hence non-perturbative techniques are needed to calculate them.
The $B_{(s)}\to \{K^{(*)},\phi\} \ell^+\ell^-$ local FFs have been computed by lattice QCD for large $q^2$ values~\cite{Bouchard:2013eph,Bailey:2015dka,Horgan:2013hoa,Horgan:2015vla}.
A lattice QCD computation for the $B \to K \ell^+\ell^-$ local FFs for low $q^2$ values has also recently appeared~\cite{Parrott:2022rgu}.
One of the main advantages of lattice QCD computations is that they have small and reducible uncertainties.
The local FFs have also been calculated using light-cone sum rules (LCSRs) at low $q^2$ and thus these two methods are complementary~\cite{Bharucha:2015bzk,Gubernari:2018wyi}.
However, the uncertainties of LCSRs are moderate ($\sim 10\%-20\%$) and not reducible below a certain threshold.
While the status of $B\to K \ell^+\ell^-$ local FFs is satisfactory given the current experimental uncertainties --- although further independent lattice QCD computations are needed --- the $B_{(s)}\to \{K^*,\phi\} \ell^+\ell^-$ local FFs definitely need improvement.
In particular, a calculation beyond the narrow width limit for the $K^*$ is necessary given the current experimental and theoretical precision~\cite{Leskovec:2024sfx}.

To combine the different theoretical calculations and to obtain the local FFs in the whole semileptonic range, it is necessary to use a parametrization.
We propose the following parametrization~\cite{Gubernari:2020eft,Gubernari:2023puw}
\begin{equation}
    \label{eq:z_exp}
    \FM{(T),\lambda}(q^2) = \frac{1}{{\cal P}_{\cal F}(q^2)\phi_{\cal F}(q^2)} \sum_{n = 0}^{\infty} \al^{\cal F}_n \, p_n(q^2)\,,
\end{equation}
where the Blaschke factors ${\cal P}_{\cal F}$, the outer functions  $\phi_{\cal F}$, and the polynomials $p_n$ are known functions.\footnote{
    Here, I do not introduce the conformal variable $z(q^2)$ for brevity.
    The details can be found in the original paper.
}
The parametrization in \refeq{z_exp} is the first one that satisfies the following two conditions simultaneously.
$(i)$ All hadronic branch cuts are taken into account --- i.e. those due to $B_s\pi(\pi)$ states --- ensuring a consistent treatment of the FFs and avoiding hard-to-quantify systematic uncertainties.
$(ii)$ The coefficients $\al^{\cal F}_n$ obey a unitarity bound:
\begin{align}
    \label{eq:bounds}
    \sum_{\cal F}
    \sum_{n = 0}^{\infty}
    \left|\al^{\cal F}_n\right|^2
     < 1\,.
\end{align}
Since, in practice, the series of~\refeq{z_exp} has to be truncated after a few terms, the unitarity bound allows us to control the truncation error in a systematic way.
We perform a simultaneous fit of all $B_{(s)}\to \{K^{(*)},\phi\} \ell^+\ell^-$ local FFs including all lattice QCD results, selected LCSRs calculations, and the unitarity bound.
This enables us to obtain all these FFs in the whole semileptonic range, with improved precision compared to the inputs used.
The results are provided as machine-readable files to facilitate their use~\cite{Gubernari:2023puw}.

The non-local FFs are clearly more complicated objects than the local FFs.
Despite preliminary studies, a lattice QCD computation of $\HM{\lambda}$ is currently out of reach~\cite{Nakayama:2020hhu}.
We then use a light-cone operator product expansion (LCOPE)~\cite{Khodjamirian:2010vf} to calculate $\HM{\lambda}$ for $q^2 \ll 4m_c^2$.
This LCOPE can be written as
\begin{align}
    \HM{\lambda}\!(q^2) = C_{\cal F}(q^2) \FM{(T),\lambda}\!(q^2) 
    + \frac{m_b \Lambda_{\rm had}}{4m_c^2 - q^2} \tilde{C}_{\cal F}(q^2) \VM{\lambda}\!(q^2)
    + {\cal O} \left(\left(\frac{m_b \Lambda_{\rm had}}{4m_c^2 - q^2}\right)^2\right)
    \,.
\end{align}
The Wilson coefficients $C_{\cal F}$ and $\tilde{C}_{\cal F}$ can be calculated perturbatively.
The $\alpha_s$ corrections to $C_{\cal F}$ are known.
On the contrary, the FFs $\FM{(T),\lambda}$ and $\VM{\lambda}$ have to be obtained by non-perturbative methods.
The $\FM{(T),\lambda}$ have already been discussed above.
While it may seem awkward that the non-local FFs depend on the local FFs (at the leading power in the LCOPE), it is somewhat intuitive that the local limit of the non-local FFs must depend on the local FFs.
The $\VM{\lambda}$ are manifestly non-local objects, and they have been calculated using LCSRs~\cite{Khodjamirian:2010vf,Gubernari:2022hxn}.
The most recent calculation for $\VM{\lambda}$ have found these contributions to be small.
In addition, it is also possible to use the experimental measurements for $B_{(s)}\to \{K^{(*)},\phi\} J/\psi$ decays, as these decays only receive a non-local FFs contribution.
For instance, for $B \to K J/\psi$ one has~\cite{Gubernari:2022hxn}
\begin{align}
    {\cal A}(B \to K J/\psi) \propto \underset{q^2\to M_\psi^2}{\text{Res}} \HM[BK]{}(q^2)
    \,.
\end{align}

To fit the LCOPE calculations and the $B_{(s)}\to \{K^{(*)},\phi\} J/\psi$, we use an analogous parametrization to that in \refeq{z_exp}:
\begin{equation}
    \label{eq:z_exp2}
    \HM{\lambda}(q^2) = \frac{1}{{\cal P}_{\cal H}(q^2)\phi_{\cal H}(q^2)} \sum_{n = 0}^{\infty} \beta^{\cal H}_n \, p_n(q^2)\,.
\end{equation}
As before, the advantage of this parametrization is that the coefficients are bounded by unitarity:
\begin{align}
    \sum_{\cal H}
    \sum_{n = 0}^{\infty}
    \left|\beta^{\cal H}_n\right|^2
     < 1\,.
\end{align}
This is the first unitarity bound for non-local FFs~\cite{Gubernari:2020eft}.
By performing this fit we obtain numerical results for the $\HM{\lambda}(q^2)$ below the open-charm threshold.
The results are provided as machine-readable files~\cite{Gubernari:2022hxn}.

\section{SM predictions and comparison with measurements}
\label{sec:SMpred}

\begin{figure}[t!]
    \centering
    \includegraphics[width=.35\textwidth]{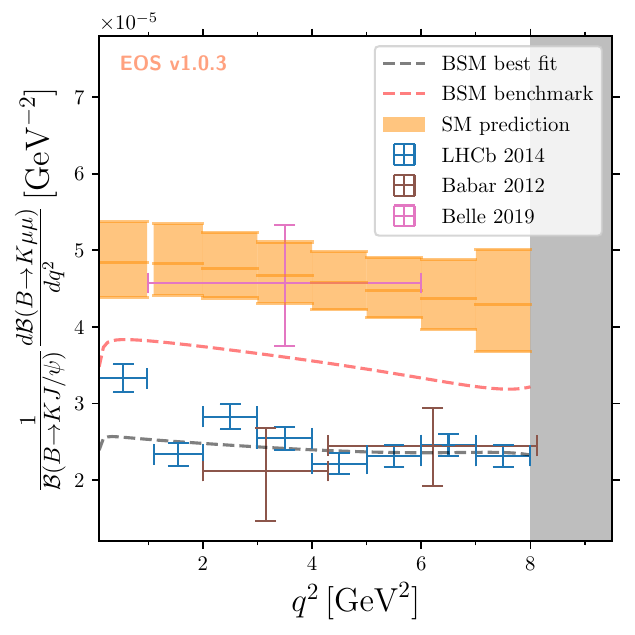} \hspace*{1cm}
    \includegraphics[width=.35\textwidth,height=.35\textwidth]{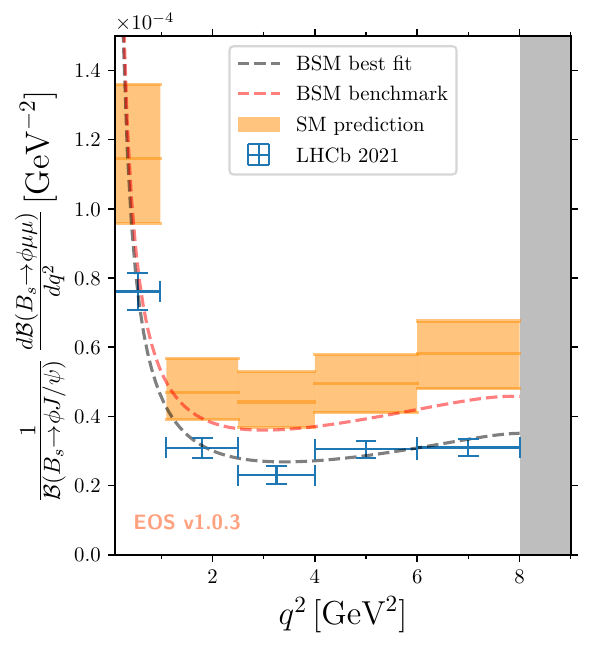} 
    \caption{%
        SM predictions for the normalised differential branching ratios taken from ref.~[9].
    }
    \label{fig:SM_Plots}
\end{figure}

Using our results for the local FFs $\FM{\lambda}$ and the non-local FFs $\HM{\lambda}$, we can obtain the SM predictions for the branching ratios and angular observables in $B_{(s)}\to \{K^*,\phi\} \mu^+\mu^-$ decays.
Two of these predictions are illustrated \reffig{SM_Plots}.
The main source of uncertainty in our predictions is the local FFs.
Still, progress on calculations for non-local FFs is urgently needed to fully control the systematic uncertainties and to clarify whether the $b\to s \mu^+ \mu^-$ anomalies are due to NP or to underestimated theoretical uncertainties.

It is also important to stress that the shifts found in $B_{(s)}\to \{K^*,\phi\} \mu^+\mu^-$ decays are coherent and $q^2$ independent.
Coherent means that the difference between SM predictions and data is approximately the same for different processes.
$q^2$ independent means that for a given observable this difference does not depend on the value of $q^2$. (This has also been elegantly pointed out in a recent paper~\cite{Bordone:2024hui}).
These two observations are crucial because, in general, a NP contribution from a heavy mediator should be both coherent and $q^2$ independent, whereas QCD effects are expected to show both a process and a $q^2$ dependence.
However, the process and $q^2$ dependence of the current shifts could be hidden in the current uncertainties, that is why progress on the theory side is crucial.

To quantify the impact of NP in each of the three processes considered, we allow for a (real) NP contribution in $C_9$ and $C_{10}$: $C_{9,10}=C_{9,10}^{SM}+C_{9,10}^{NP}$.
The largest SM pull is in $B \to K \mu^+\mu^-$ ($5.7\sigma$), since the predictions for this channel are very precise~\cite{Gubernari:2022hxn}. 
Nevertheless, the three processes show coherent $68\%$ intervals for $C_{9,10}^{NP}$, as anticipated above.

\section{Aside: $\boldsymbol{\Lambda_b\to\Lambda \ell^+\ell^-}$ decays}
\label{sec:baryon}

\begin{figure}[t!bp]
	\centering
	\includegraphics[width=0.85\textwidth]{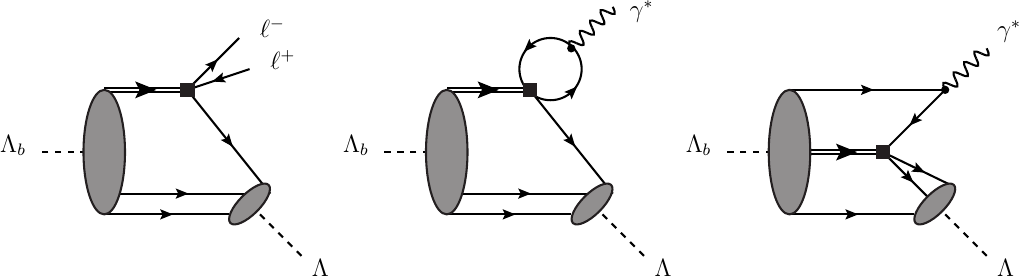}
	\caption{
            The local (left), non-local charm-loop (center), and non-local penguin (right) contributions to \hbox{$\Lambda_b \to \Lambda \ell^+\ell^-$} decays.
            Image taken from ref.~[25].
    }
	\label{fig:Lambdab}
\end{figure}

The $\Lambda_b \to \Lambda \ell^+\ell^-$ decays offer the possibility to probe $b \to s \ell^+\ell^-$ transitions using baryon decays.
If the $b\to s \mu^+ \mu^-$ anomalies are due to NP, the same discrepancy between SM predictions and data will appear in $\Lambda_b \to \Lambda \mu^+\mu^-$ decays.
In contrast, the non-local FFs are expected to differ between the mesonic and baryonic cases.
This is mainly due to the different spin of the hadrons and the different number of valence quarks within them.

In order to carry out this test, a significant improvement is needed on both the theoretical and experimental sides.
The uncertainties of the current measurements are statistically dominated, therefore future measurements will be more precise.
On the theoretical side~\cite{Detmold:2016pkz,Blake:2022vfl,Meinel:2023wyg}, there is a need to improve and develop new techniques to improve the accuracy of the non-local FFs in particular.
For instance, there is no estimate of the non-local charm loop contributions beyond naive factorisation (central panel of \reffig{Lambdab}).
A step in this direction has been taken in one of my recent publications~\cite{Feldmann:2023plv}.
Here the contribution of the penguin operators has been calculated for the first time (right panel of \reffig{Lambdab}).
These contributions are analogous to the annihilation topologies in mesonic decays.
We find that these contributions are small and hence the systematic uncertainties coming from the penguin operators can be considered to be under control.

\section{Summary and conclusion}
\label{sec:conc}

The $b$-hadron decays mediated by $b \to s \ell^+\ell^-$ transitions are undoubtedly a very active and fascinating area of research.
Both the experimental and the theoretical communities are putting a great deal of effort into reducing the current uncertainties.
An improvement of the experimental uncertainties can be anticipated, as most of the measurements are statistically limited.
Theoretical uncertainties are also expected to improve, especially those due to local form factors, through new and more advanced lattice QCD calculations.
The predictions for non-local form factors are way more challenging to improve, and they are beyond the reach of present lattice QCD techniques.
Nevertheless, there is a lot of room for improvement, especially in the $\Lambda_b \to \Lambda \ell^+\ell^-$ decays.
For instance, combined unitarity analyses, improvements in the $b$ hadron distribution amplitudes, and higher order calculations will certainly help to reduce the uncertainties.
In conclusion, although the road to improving SM predictions for $b \to s \ell^+\ell^-$ decays may seem daunting, we must have the conviction that with dedication and innovation even the most complex calculations can be mastered.

\section*{Acknowledgements}

I would like to thank all the organisers of the ``58$^{\rm th}$ Rencontres de Moriond'', not only for inviting me to present my work, but also for the excellent and smooth planning of every detail of the conference. 
It is certainly a conference with a unique atmosphere that encourages exchange between researchers.
I would also like to thank M. Reboud for his feedback on this document and for pointing out (embarrassing) typos.
\sloppy This work has been partially supported by STFC consolidated grants ST/T000694/1 and ST/X000664/1.


\section*{References}

\begin{thebibliography}{99}

%\cite{Allanach:2023uxz}
\bibitem{Allanach:2023uxz}
B.~Allanach and A.~Mullin,
%``Plan B: new Z' models for b \textrightarrow{} s\ensuremath{\ell}$^{+}$\ensuremath{\ell}$^{−}$ anomalies,''
JHEP \textbf{09} (2023), 173,
%doi:10.1007/JHEP09(2023)173
arXiv:2306.08669.
%11 citations counted in INSPIRE as of 01 Apr 2024

%\cite{Crivellin:2023saq}
\bibitem{Crivellin:2023saq}
A.~Crivellin and M.~Kirk,
%``Diquark explanation of b\textrightarrow{}s\ensuremath{\ell}+\ensuremath{\ell}-,''
Phys. Rev. D \textbf{108} (2023) no.11, 11,
%doi:10.1103/PhysRevD.108.L111701
arXiv:2309.07205.
%4 citations counted in INSPIRE as of 01 Apr 2024

%\cite{LHCb:2014cxe}
\bibitem{LHCb:2014cxe}
R.~Aaij \textit{et al.} [LHCb],
%``Differential branching fractions and isospin asymmetries of $B \to K^{(*)} \mu^+ \mu^-$ decays,''
JHEP \textbf{06} (2014), 133,
%doi:10.1007/JHEP06(2014)133
arXiv:1403.8044.
%615 citations counted in INSPIRE as of 01 Apr 2024

%\cite{CMS:2024syx}
\bibitem{CMS:2024syx}
A.~Hayrapetyan \textit{et al.} [CMS],
%``Test of lepton flavor universality in B$^{\pm}$$\to$ K$^{\pm}\mu^+\mu^-$ and B$^{\pm}$$\to$ K$^{\pm}$e$^+$e$^-$ decays in proton-proton collisions at $\sqrt{s}$ = 13 TeV,''
arXiv:2401.07090.
%4 citations counted in INSPIRE as of 01 Apr 2024

%\cite{LHCb:2016ykl}
\bibitem{LHCb:2016ykl}
R.~Aaij \textit{et al.} [LHCb],
%``Measurements of the S-wave fraction in $B^{0}\rightarrow K^{+}\pi^{-}\mu^{+}\mu^{-}$ decays and the $B^{0}\rightarrow K^{\ast}(892)^{0}\mu^{+}\mu^{-}$ differential branching fraction,''
JHEP \textbf{11} (2016), 047,
%[erratum: JHEP \textbf{04} (2017), 142]
%doi:10.1007/JHEP11(2016)047
arXiv:1606.04731.
%301 citations counted in INSPIRE as of 01 Apr 2024

%\cite{LHCb:2020lmf}
\bibitem{LHCb:2020lmf}
R.~Aaij \textit{et al.} [LHCb],
%``Measurement of $CP$-Averaged Observables in the $B^{0}\rightarrow K^{*0}\mu^{+}\mu^{-}$ Decay,''
Phys. Rev. Lett. \textbf{125} (2020) no.1, 011802,
%doi:10.1103/PhysRevLett.125.011802
arXiv:2003.04831.
%364 citations counted in INSPIRE as of 01 Apr 2024

%\cite{LHCb:2021zwz}
\bibitem{LHCb:2021zwz}
R.~Aaij \textit{et al.} [LHCb],
%``Branching Fraction Measurements of the Rare $B^0_s\rightarrow\phi\mu^+\mu^-$ and $B^0_s\rightarrow f_2^\prime(1525)\mu^+\mu^-$- Decays,''
Phys. Rev. Lett. \textbf{127} (2021) no.15, 151801,
%doi:10.1103/PhysRevLett.127.151801
arXiv:2105.14007.
%186 citations counted in INSPIRE as of 01 Apr 2024

%\cite{Parrott:2022zte}
\bibitem{Parrott:2022zte}
W.~G.~Parrott \textit{et al.} [HPQCD],
%``Standard Model predictions for B\textrightarrow{}K\ensuremath{\ell}+\ensuremath{\ell}-, B\textrightarrow{}K\ensuremath{\ell}1-\ensuremath{\ell}2+ and B\textrightarrow{}K\ensuremath{\nu}\ensuremath{\nu}\textasciimacron{} using form factors from Nf=2+1+1 lattice QCD,''
Phys. Rev. D \textbf{107} (2023) no.1, 014511,
%[erratum: Phys. Rev. D \textbf{107} (2023) no.11, 119903]
%doi:10.1103/PhysRevD.107.014511
arXiv:2207.13371.
%48 citations counted in INSPIRE as of 01 Apr 2024

%\cite{Gubernari:2022hxn}
\bibitem{Gubernari:2022hxn}
N.~Gubernari \textit{et al.},
%``Improved theory predictions and global analysis of exclusive $b \to s\mu^+\mu^-$ processes,''
JHEP \textbf{09} (2022), 133,
%doi:10.1007/JHEP09(2022)133
arXiv:2206.03797.
%80 citations counted in INSPIRE as of 01 Apr 2024

%\cite{LHCb:2022qnv}
\bibitem{LHCb:2022qnv}
R.~Aaij \textit{et al.} [LHCb],
%``Test of lepton universality in $b \rightarrow s \ell^+ \ell^-$ decays,''
Phys. Rev. Lett. \textbf{131} (2023) no.5, 051803,
%doi:10.1103/PhysRevLett.131.051803
arXiv:2212.09152.
%172 citations counted in INSPIRE as of 01 Apr 2024

%\cite{Buchalla:1995vs}
\bibitem{Buchalla:1995vs}
G.~Buchalla \textit{et al.},
%``Weak decays beyond leading logarithms,''
Rev. Mod. Phys. \textbf{68} (1996), 1125-1144,
%doi:10.1103/RevModPhys.68.1125
arXiv:hep-ph/9512380.
%2984 citations counted in INSPIRE as of 02 Apr 2024

%\cite{Gubernari:2020eft}
\bibitem{Gubernari:2020eft}
N.~Gubernari, D.~van Dyk and J.~Virto,
%``Non-local matrix elements in $B_{(s)}\to \{K^{(*)},\phi\}\ell^+\ell^-$,''
JHEP \textbf{02} (2021), 088,
%doi:10.1007/JHEP02(2021)088
arXiv:2011.09813.
%104 citations counted in INSPIRE as of 02 Apr 2024

%\cite{Bouchard:2013eph}
\bibitem{Bouchard:2013eph}
C.~Bouchard \textit{et al.} [HPQCD],
%``Rare decay $B \to K \ell^+ \ell^-$ form factors from lattice QCD,''
Phys. Rev. D \textbf{88} (2013) no.5, 054509,
%doi:10.1103/PhysRevD.88.054509
arXiv:1306.2384.
%173 citations counted in INSPIRE as of 02 Apr 2024

%\cite{Bailey:2015dka}
\bibitem{Bailey:2015dka}
J.~A.~Bailey \textit{et al.},
%``$B\to Kl^+l^-$ Decay Form Factors from Three-Flavor Lattice QCD,''
Phys. Rev. D \textbf{93} (2016) no.2, 025026,
%doi:10.1103/PhysRevD.93.025026
arXiv:1509.06235.
%158 citations counted in INSPIRE as of 02 Apr 2024

%\cite{Horgan:2013hoa}
\bibitem{Horgan:2013hoa}
R.~R.~Horgan \textit{et al.},
%``Lattice QCD calculation of form factors describing the rare decays $B \to K^* \ell^+ \ell^-$ and $B_s \to \phi \ell^+ \ell^-$,''
Phys. Rev. D \textbf{89} (2014) no.9, 094501,
%doi:10.1103/PhysRevD.89.094501
arXiv:1310.3722.
%253 citations counted in INSPIRE as of 02 Apr 2024

%\cite{Horgan:2015vla}
\bibitem{Horgan:2015vla}
R.~R.~Horgan \textit{et al.},
%``Rare $B$ decays using lattice QCD form factors,''
PoS LATTICE2014 (2015), 372,
%doi:10.22323/1.214.0372
arXiv:1501.00367.
%124 citations counted in INSPIRE as of 02 Apr 2024

%\cite{Parrott:2022rgu}
\bibitem{Parrott:2022rgu}
W.~G.~Parrott \textit{et al.} [HPQCD],
%``B\textrightarrow{}K and D\textrightarrow{}K form factors from fully relativistic lattice QCD,''
Phys. Rev. D \textbf{107} (2023) no.1, 014510,
%doi:10.1103/PhysRevD.107.014510
arXiv:2207.12468.
%57 citations counted in INSPIRE as of 02 Apr 2024

%\cite{Bharucha:2015bzk}
\bibitem{Bharucha:2015bzk}
A.~Bharucha, D.~M.~Straub and R.~Zwicky,
%``$B\to V\ell^+\ell^-$ in the Standard Model from light-cone sum rules,''
JHEP \textbf{08} (2016), 098,
%doi:10.1007/JHEP08(2016)098
arXiv:1503.05534.
%566 citations counted in INSPIRE as of 02 Apr 2024

%\cite{Gubernari:2018wyi}
\bibitem{Gubernari:2018wyi}
N.~Gubernari, A.~Kokulu and D.~van Dyk,
%``$B\to P$ and $B\to V$ Form Factors from $B$-Meson Light-Cone Sum Rules beyond Leading Twist,''
JHEP \textbf{01} (2019), 150,
%doi:10.1007/JHEP01(2019)150
arXiv:1811.00983.
%199 citations counted in INSPIRE as of 02 Apr 2024

%\cite{Leskovec:2024sfx}
\bibitem{Leskovec:2024sfx}
L.~Leskovec \textit{et al.},
%``Lattice outlook on $B\to\rho\ell\bar{\nu}$ and $B\to K^\star \ell \ell$,''
contribution to CKM 2023,
arXiv:2403.19543.
%0 citations counted in INSPIRE as of 03 Apr 2024

%\cite{Gubernari:2023puw}
\bibitem{Gubernari:2023puw}
N.~Gubernari \textit{et al.},
%``Dispersive analysis of B \textrightarrow{} K$^{(*)}$ and B$_{s}$\textrightarrow{} \ensuremath{\phi} form factors,''
JHEP \textbf{12} (2023), 153,
%doi:10.1007/JHEP12(2023)153
arXiv:2305.06301.
%22 citations counted in INSPIRE as of 03 Apr 2024

%\cite{Nakayama:2020hhu}
\bibitem{Nakayama:2020hhu}
K.~Nakayama \textit{et al.} [JLQCD],
%``Charmonium contribution to $B \to K\ell^+\ell^-$: testing the factorization approximation on the lattice,''
PoS LATTICE2019 (2020), 062,
%doi:10.22323/1.363.0062
arXiv:2001.10911.
%5 citations counted in INSPIRE as of 03 Apr 2024

%\cite{Khodjamirian:2010vf}
\bibitem{Khodjamirian:2010vf}
A.~Khodjamirian \textit{et al.},
%``Charm-loop effect in $B \to K^{(*)} \ell^{+} \ell^{-}$ and $B\to K^*\gamma$,''
JHEP \textbf{09} (2010), 089,
%doi:10.1007/JHEP09(2010)089
arXiv:1006.4945.
%486 citations counted in INSPIRE as of 03 Apr 2024

%\cite{Bordone:2024hui}
\bibitem{Bordone:2024hui}
M.~Bordone, G.~isidori, S.~M\"achler and A.~Tinari,
%``Short- vs. long-distance physics in $B\to K^{(*)} \ell^+\ell^-$: a data-driven analysis,''
arXiv:2401.18007.
%3 citations counted in INSPIRE as of 03 Apr 2024

%\cite{Detmold:2016pkz}
\bibitem{Detmold:2016pkz}
W.~Detmold and S.~Meinel,
%``$\Lambda_b \to \Lambda \ell^+ \ell^-$ form factors, differential branching fraction, and angular observables from lattice QCD with relativistic $b$ quarks,''
Phys. Rev. D \textbf{93} (2016) no.7, 074501,
%doi:10.1103/PhysRevD.93.074501
arXiv:1602.01399.
%131 citations counted in INSPIRE as of 05 Apr 2024

%\cite{Meinel:2023wyg}
\bibitem{Meinel:2023wyg}
S.~Meinel,
%``Status of next-generation $\Lambda_b \to p, \Lambda, \Lambda_c$ form-factor calculations,''
PoS LATTICE2023 (2024), 275
%doi:10.22323/1.453.0275
arXiv:2309.01821.
%4 citations counted in INSPIRE as of 05 Apr 2024

%\cite{Blake:2022vfl}
\bibitem{Blake:2022vfl}
T.~Blake \textit{et al.},
%``Dispersive bounds for local form factors in \ensuremath{\Lambda}b\textrightarrow{}\ensuremath{\Lambda} transitions,''
Phys. Rev. D \textbf{108} (2023) no.9, 094509,
%doi:10.1103/PhysRevD.108.094509
arXiv:2205.06041.
%15 citations counted in INSPIRE as of 05 Apr 2024





%\cite{Feldmann:2023plv}
\bibitem{Feldmann:2023plv}
T.~Feldmann and N.~Gubernari,
%``Non-factorisable contributions of strong-penguin operators in \ensuremath{\Lambda}$_{b}$\textrightarrow{} \ensuremath{\Lambda}\ensuremath{\ell}$^{+}$\ensuremath{\ell}$^{−}$ decays,''
JHEP \textbf{03} (2024), 152,
%doi:10.1007/JHEP03(2024)152
arXiv:2312.14146.
%1 citations counted in INSPIRE as of 01 Apr 2024





\end{thebibliography}
\end{document}